\documentclass[a4paper,floatfix,rmp,twocolumn,showkeys,superscriptaddress]{revtex4}
\usepackage[latin1]{inputenc}
\usepackage{natbib}
\usepackage{graphicx}

\begin{document}

\title{Modularity ``for free'' in genome architecture?}

\providecommand{\ICREA}
{ICREA-Complex Systems Lab, Universitat Pompeu Fabra (GRIB),%
 Dr Aiguader 80, 08003 Barcelona, Spain}
\providecommand{\SFI}
{Santa Fe Institute, 1399 Hyde Park Road, Santa Fe NM 87501, USA}

\author{Ricard V. Solé}
\affiliation{\ICREA}
\affiliation{\SFI}
\author{Pau Fernández}
\affiliation{\ICREA}

\vspace{0.4cm}
\begin{abstract}
{\large \bf Abstract} \vspace{0.2cm} \\ 

{\bf  Background}  \hspace{.2em}   Recent  models  of  genome-proteome
evolution  have shown that  some of  the key  traits displayed  by the
global structure of  cellular networks might be a  natural result of a
duplication-diversification (DD)  process. One of  the consequences of
such evolution is the emergence of a small world architecture together
with a scale-free distribution of interactions.  Here we show that the
domain of parameter space were  such structure emerges is related to a
phase  transition  phenomenon.   At  this  transition  point,  modular
architecture spontaneously  emerges as a byproduct of  the DD process.
\vspace{0.15cm} \\

{\bf  Results}   \hspace{.2em}  Although   the  DD  models   lack  any
functionality and  are thus free from  meeting functional constraints,
they show  the observed features  displayed by the real  proteome maps
when  tuned close  to a  sharp  transition point  separating a  highly
connected graph  from a disconnected system.  Close  to such boundary,
the maps  are shown  to display scale-free  hierarchical organization,
behave as small worlds and exhibit modularity.  \vspace{0.15cm} \\

{\bf  Conclusions}  \hspace{.2em}   It  is  conjectured  that  natural
selection  tuned the  average  connectivity  in such  a  way that  the
network  reaches a sparse  graph of  connections.  One  consequence of
such scenario is  that the scaling laws and  the essential ingredients
for  building  a modular  net  emerge {\em  for  free}  close to  such
transition.
\end{abstract}


\maketitle

\section{Introduction}

The intimate structure  of cellular life is largely  associated to the
networks  of interactions  among  different types  of molecules.   The
structure of  cellular networks, from  the genome and the  proteome to
the   metabolome  strongly   departs  from   a  simple   random  graph
\cite{SoleSatorras2002}.   Instead,   these  nets  display   a  highly
heterogeneous   architecture:   most   units   (genes,   proteins   or
metabolites)  are linked to  a few  other units  but invariably  a few
units exhibit  a large number  of links.  Such heterogeneity  has been
also  found in a  wide spectrum  of complex  systems, from  natural to
artificial  \cite{BornholdtBook}.  More  importantly,  the topological
organization  of complex  nets might  pervade their  efficiency, their
robustness and their fragility under perturbations \cite{Albert2000}.

\begin{figure}  
{\centering 
\includegraphics[width=8cm]{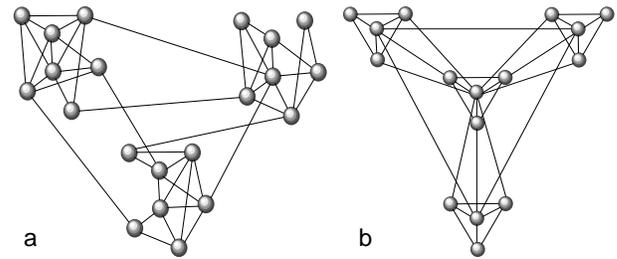}}
  \caption{
    \label{modular1}
Two  examples of  modular networks.  In  (a) three  basic modules  are
involved, each one  involving a set of nodes  randomly connected among
them with  some given  probability.  Each node  can also  be connected
(with a smaller probability) with other nodes in other modules. In (b)
a hierarchical  network is shown,  created by repeating a  given basic
motif at different levels \cite{Ravasz2002}.}
\end{figure}

The analysis of network structure  and dynamics offers a new window to
answer     questions    relating     evolution     of    biocomplexity
\cite{SoleCOMPLEXITYNETS}.    Networks   are   organized   in   highly
non-random ways and the topological organization of their connectivity
allow     to     quantitatively     define     some     characteristic
traits.  Understanding  the origins  of  such  properties requires  an
understanding  of  the  evolutionary  mechanisms that  generate  these
networks. Since properly defined  quantitative traits can be measured,
models  are strongly constrained  to reproduce  a well-defined  set of
features.

From  a  statistical  point  of  view,  protein-protein  or  gene-gene
interaction    maps   can    be   viewed    as   a    random   network
\cite{BollobasBook,BornholdtBook},   in   which   the   vertices
represent  the  proteins (genes)  and  an  edge  between two  vertices
indicates  the  presence  of  an interaction  between  the  respective
proteins.  In  this paper  we restrict our  analysis to  an undirected
graph of protein-protein interactions, but some of our conclusions can
be translated to regulatory networks.

Mathematically,   the   proteome   graph   is  defined   by   a   pair
$\Omega_p=(W_p, E_p)$, where $W_p=\{ p_i \}, (i=1, ..., N)$ is the set
of  $N$  proteins  and  $E_p=\{   \{p_i,  p_j\}  \}$  is  the  set  of
edges/connections  between  proteins.    The  {\em  adjacency  matrix}
$\xi_{ij}$ indicates that an interaction exists between proteins $p_i,
p_j \in \Omega_p$  ($\xi_{ij} = 1$) or that  the interaction is absent
($\xi_{ij}  =  0$).   Two  connected  proteins are  thus  called  {\em
adjacent} and  the {\em degree}  of a given  protein is the  number of
edges that connect it with other proteins.

The analysis of metabolic  pathways, protein interaction maps, genetic
regulatory  networks and  gene expression  data reveals  that cellular
webs belong to a class of network topologies known as {\em scale-free}
(SF) networks \cite{Jeong2000,Fell2000}.  A SF net is characterized by
a so-called degree  distribution $P(k)$ displaying power-law behavior.
Here $P(k)$  is the probability of  finding a unit which  is linked to
$k$ other units  and typically decays as $P(k)  \sim k^{-\gamma}$ with
$2<\gamma<3$.    Here   links  correspond,   for   protein  maps,   to
protein-protein  interactions. These networks  are also  small worlds:
the average  number of steps  $d$ required in  order to jump  from one
protein to another through the network is very small \cite{Fell2000}.

Scale-free graphs  have been shown  to emerge from different  types of
mechanisms
\cite{Albert2001,DorogovtsevBook,Newman2003,Caldarelli2003}.   Most of
them involve (explicitly or implicitly) a multiplicative process known
as {\em preferential attachment} \cite{Barabasi1999}.  In its standard
form, it relies  on a popularity principle (rich  gets richer): as new
node are  added to the system,  they tend to  attach preferentially to
nodes  with higher  degree, in  the Barabasi-Albert  (BA)  model, this
process  leads   to  a  SF   distribution  with  $P(k)   \sim  k^{-3}$
\cite{Barabasi1999}.

An  additional feature is  the presence  of modular  architecture with
well-defined hierarchical properties \cite{Ravasz2002}.  An example of
modular network is shown in figure \ref{modular1}a. Here three sets of
nodes appear  more connected among them  than with other  nodes in the
graph. Three modules are thus  naturally defined (although only from a
topological  point of  view).  In this  particular  model (defined  in
\cite{Ravasz2002}) nodes  inside each  module are randomly  wired with
some probability $p$, as in so-called Erd\"os-Renyi (ER) graphs.  They
are also  linked to nodes in  other modules with  a probability $q<p$.
Such  networks  exhibit a  Poissonian  degree distribution.   Cellular
networks,  however, are  not  poissonian, but  are certainly  modular,
exhibiting hierarchical organization \cite{Ravasz2002}.

In  table I we  summarize the  differences between  networks generated
through  random  wiring (ER),  preferential  attachment  (from the  BA
model) and the actual proteome map. It is interesting that none of the
models gives modular  architecture \cite{Ravasz2002}. Protein modules,
for  example, result from  the binding  of multiple  protein molecules
forming stable complexes.  The  presence of hierarchies has been shown
to be measurable from the so called {\em clustering coefficient} $C_i$
which  measures the  fraction  of  neighbors of  this  node that  are
neighbours among them, i. e.
\begin{equation}
C_{i}=\frac{2E_{i}}{k_{i}(k_{i}-1)}
\end{equation}
In this formula, $E_{i}$ is  the number of links between neighbours of
$i$ (with degree) $k_i$, $k_i(k_i-1)$  being thus the total number of
possible links between  neighbours of $i$. The average  of $C_i$, that
is, $C(N) =  \sum_i C_i /N$, describes in  general the {\em clustering
coefficient} of a  network. This measure has been  observed to be much
higher in real networks than for  random graphs in a variety of fields
\cite{DorogovtsevBook}, and  in particular, it has also  been shown to
display a scale-free distribution \cite{Ravasz2002}.

It is generally acknowledged  that modules define functional units and
as       such       are       the      target       of       selection
\cite{Hartwell1999,Sole2002_PHYSA}.  In  this  context,  some  authors
suggested that {\em general ``design principles' -profoundly shaped by
the  constraints of  evolution-govern  the structure  and function  of
modules''}           \cite{Hartwell1999}           (see           also
\cite{WagnerG1996b,vonDassow1999}).

Modules have been found in biological systems at multiple levels, from
RNA   structures  \cite{Ancel2000}   to  the   cerebral   cortex  (see
\cite{SoleCOMPLEXITYNETS}  and  references  therein). This  widespread
character of modular  organization pervades the functional association
between  compartmentalization   and  evolution.   Modules   have  been
variously  defined  as  functionally buffered,  robust,  independently
controlled,   plastic  in   composition   and  interconnectivity   and
evolutionarily conserved. The  evolutionary conservation of modules is
clearly  appreciated in  gene networks  involved in  early development
\cite{vonDassow2000,  Munro,  SoleTREE}.   The  argument is  that  the
special features of some of  these modules are tightly linked to their
robustness under different sources of noise.

The  modular character  of  biological  networks is  assumed  to be  a
consequence    of    both    their   robustness    and    evolvability
\cite{WagnerG1996b}. In this context, modularity would evolve through a decrease 
of pleiotropy \cite{WagnerAltenberg}. Since   they   somewhat    define   separated
compartments,  they  would act  as  buffers  against lethal  mutations
perhaps  facilitating  variation   \cite{Wolf2003}.   In  a  different
context, it  has been suggested  that modularity might arise  from the
intrinsic  structure of  the non-metric  mapping between  genotype and
phenotype \cite{Stadler2001}. Although  functionality must pervade the
selection  of some  modular  structures, here  we  show, by  exploring
available  data   and  simple  models  of   proteome  evolution,  that
proto-modules  might  actually  result from  a  duplication-divergence
process without  any predefined functional meaning.   If correct, this
observation would  actually indicate that modular  structures would be
already in place as a byproduct of genome growth.

\begin{table}[tp]
\begin{tabular}{|c|c|c|c|} \hline
\bf Property  & \bf ER  graph& \bf BA  model & \bf proteome  \\ \hline
 $C(N)$  & $N^{-1}$  & $N^{-1/2}$  &  independent \\  \hline $C(k)$  &
 independent &  $k^{-1}$ &  $k^{-1}$ \\ \hline  $P(k)$ &  Poissonian &
 $k^{-\gamma}$ & $(k+k_0)^{-\gamma}e^{-k/k_c})$ \\ \hline Modules & no
 & no & yes \\ \hline
\end{tabular}
\caption{
  Global properties  displayed by different types  of graphs, compared
with those  exhibited by a  hierarchical system, such as  the proteome
map. Here the scaling exponent is $2<\gamma<3$.}
\end{table}

\section{Results and Discussion}

\subsection{Phase transition in the proteome evolution model}

Any  model  involving genome  evolution  must  take  into account  the
leading mechanism that  appears to be responsible of  its growth: gene
duplication.  Through  gene duplication \cite{OhonoBook}  new elements
are incorporated  to the system,  initially introducing an  element of
redundancy, since genes are duplicated and thus their connections with
others  too. Afterwards,  divergence or  loss of  function  occurs and
either new functions/interactions are developed or pseudogenes (i. e.
nonfunctional copies of duplicated genes) generated.

\begin{figure*}
{\centering \includegraphics[width=11cm]{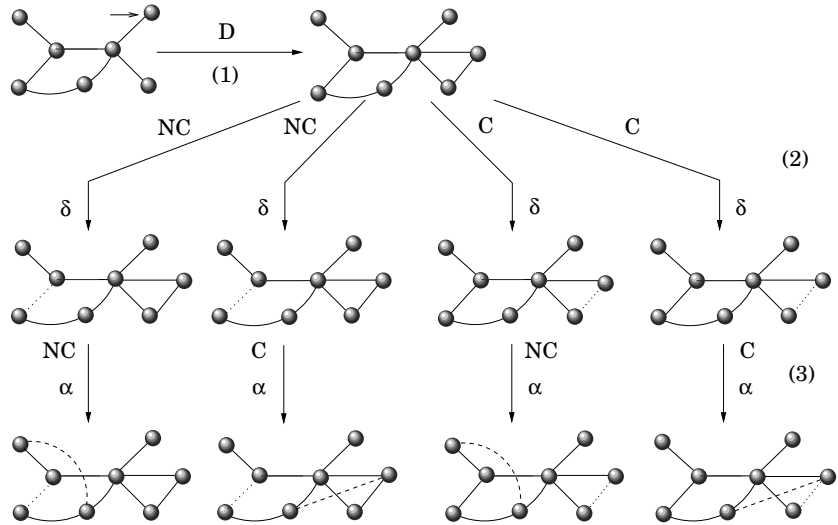}}
  \caption{
    \label{rules}
    Rules of  proteome growth in the four  possible scenarios.  First,
    (1) duplication  occurs  after randomly  selecting  a node  (small
    arrow).  Then (2) deletion  of connections occurs with probability
    $\delta$.  This event can be correlated (C) when the deleted links
    are connected  to the newly  generated node or  uncorrelated (NC),
    when  all  links are  considered  for  deletion.  Finally (3)  new
    connections are  generated with  probability $\alpha$, again  in a
    correlated or uncorrelated way. The time scales at which different
    events  occur are known  to be  very different:  duplication takes
    place  at a  much  slower rate,  whereas  rewiring is much  faster.
    Additionally, the  specific rates at which each  event occur might
    involve   preferential   attachment    to   proteins   of   higher
    connectivities. All these variants can be included. }
\end{figure*}

In trying to understand  genome evolution under a network perspective,
two  possible   approaches  can  be  followed.    First,  the  network
architecture  is given  and the  dynamics of  gene regulation  and its
stability can be explored by changing well-defined network parameters,
such               as               average               connectivity
\cite{Kauffman1993,Kauffman1969,Aldana2003}.    A  different  approach
would consider the  process itself of network growth.   A simple model
of this process can be  constructed by using a graph representation of
the genome,  where genes are  the nodes and  links are the  edges.  At
each  time step a  duplication event  takes place,  and the  number of
genes in the  system provides a natural time  scale, although the rate
of  link rewiring  i much  faster than  the rate  of  duplication (see
below).  Two independent studies,  involving both analytic results and
data analysis,  presented simple models of  proteome network evolution
through gene  duplication and diversification. These  models were able
to explain a large part  of the observed complexity of protein network
architecture, particularly  the presence  of small world  patterns and
the scale-free behavior. Their results  were compared with some of the
statistical   pattern   with  those   observed   from  proteome   maps
\cite{Sole2001SFI,Sole2002_ACS,Vazquez2001,Vazquez2003,SoleSatorras2002}.
Two  other  studies presented  closely  related  models using  protein
domains  as  the   basic  units  \cite{Wuchty2001,Rzhetsky2001}  again
revealing that the complex patterns found in cellular interaction maps
emerge from  these microscopic laws of genome  evolution. Further work
has  confirmed these  results  \cite{Bahn2002,Huynen2003} confirm  the
basic predictions presented in  those original papers. Further work in
this  area involves  the exploration  of  the origins  of the  protein
universe  structure,  again under  simple  models  of duplication  and
diversification  \cite{Koonin2002,Shakhnovich2002}.  Although previous
papers  have explored some  average traits  of these  interaction maps
(such as  their scale-free structure  and the presence  of small-world
architecture) here  we analyse  the patterns of  correlations emerging
from  them  and in  particular  the  presence  or absence  of  modular
organization.

The time evolution  can be described in terms of  the number of links,
i. e. we can write down a discrete equation for the link dynamics:
\begin{equation}
 L_{n+1} = L_n + \Gamma \left ( \{ K_i(n) \}, \delta, \alpha \right )  
\end{equation} 
or, using the approximation $dL_n/dn \approx L_{n+1}-L_n$, the continuous model:
\begin{equation}
 {d L_n \over dn}  = \Gamma \left (  \{ K_i(n) \}, \delta, \alpha \right )  
\end{equation} 
Using the chain rule, we have
\begin{equation}
 {d L_n \over dn} = {1 \over 2} K_n + {n \over 2}  {d K_n \over dn} 
\end{equation}
and the previous dynamical equation for links is transformed into a differential equation 
for the average degree: 
\begin{equation}
 {d K_n \over dn}  =  {n \over 2} \left [ \Gamma 
\left (  \{ K_i(n) \}, \delta, \alpha \right )  -  {1 \over 2} K_n \right ]
\end{equation} 
Here the  functional form  of $\Gamma(x)$ will  depend on  some given
(perhaps time-dependent)  parameters such as rate  of removal $\delta$
or creation  $\alpha$ of links  as well as  of the internal  state, as
defined by the  distribution of links at a  given step, here indicated
as $\{ K_i(n) \}$ (with $i=1, ..., n$).

Different functional forms might  be chosen, including rates of change
that depend on  the degree of the node, as  suggested by some studies.
Although duplication rate would be expected to depend on the number of
links          too,          this         seems          controversial
\cite{KooninBMC2003,FraserBMC2003,BloomBMC2003}.

\begin{figure}
  {\centering \includegraphics[width=8cm]{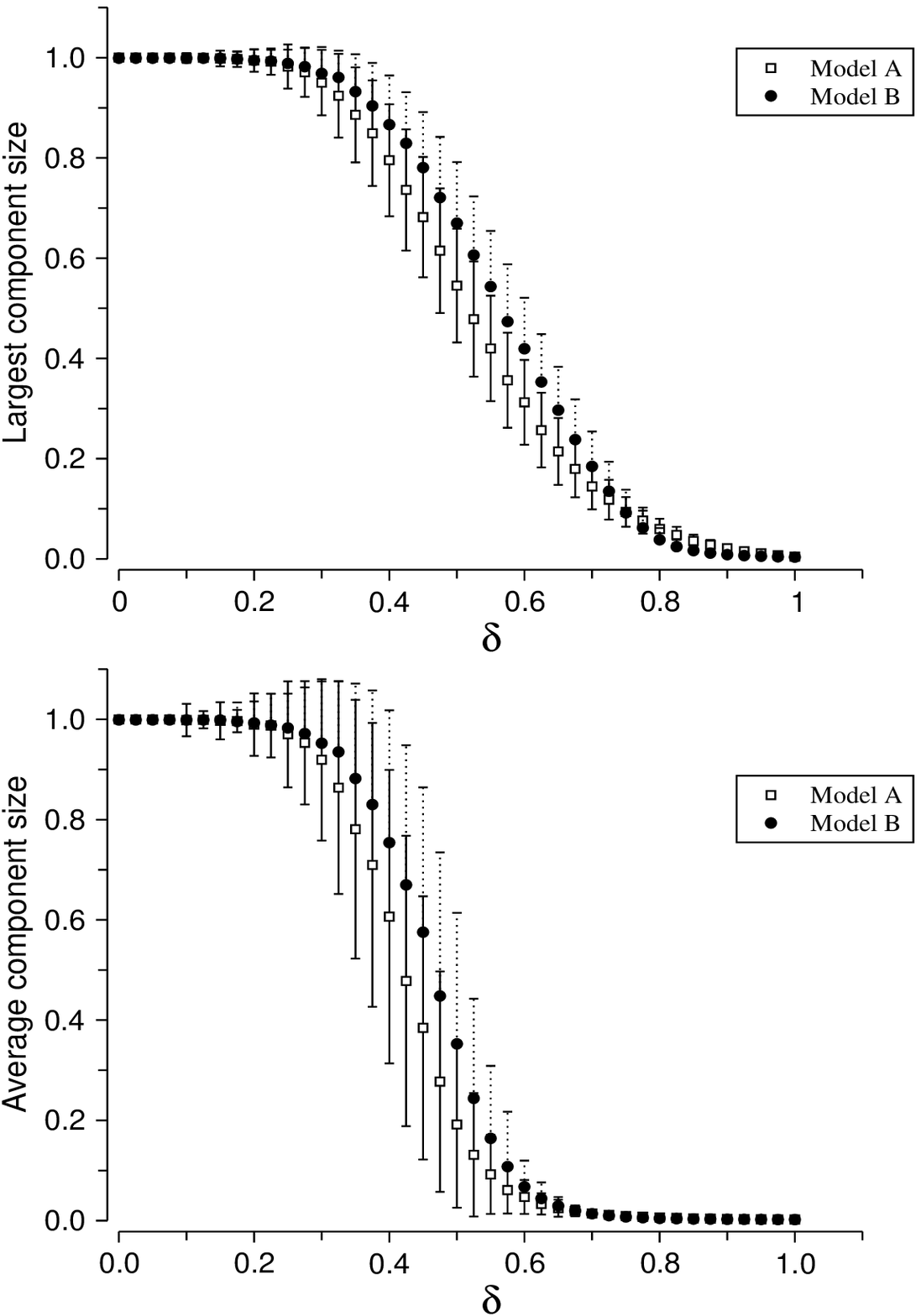}}
  \caption{
Phase  transition in  the  genome growth  models.   Here $N=10^3$  and
averages have been  performed over $R=10^3$ replicas. Here  the size of
the largest component and the  average size are shown against the rate
of link  removal $\delta$.  The  predicted phase transition  occurs at
$\delta_c  \approx  0.5$.  Due  to  the finite  (small)  size  of  our
networks, the transition appears to be less sharp than expected.}
\end{figure}

The  simplest situation would  involve pure  duplication with  no link
removal or rewiring.  This situation corresponds to $\Gamma \left ( \{
K_i(n)  \}, \delta, \alpha  \right )=K_n$  and thus  we would  have $d
K_n/dn = 2K_n/n$ with a straightforward analytic solution:
\begin{equation}
 K_n  = K_o \left (  {n \over n_o} \right )^2  
\end{equation} 
where  $n_0$ and $K_0$  are the  initial number  of links  and average
degree, respectively.  As a  consequence, an explosive increase in the
connectivity will be obtained.  Since cellular networks are sparse, we
conclude that  links have  to be deleted  at a  fast pace in  order to
reach a low, finite number of links per unit.

\begin{figure*}
  {\centering \includegraphics[width=14cm]{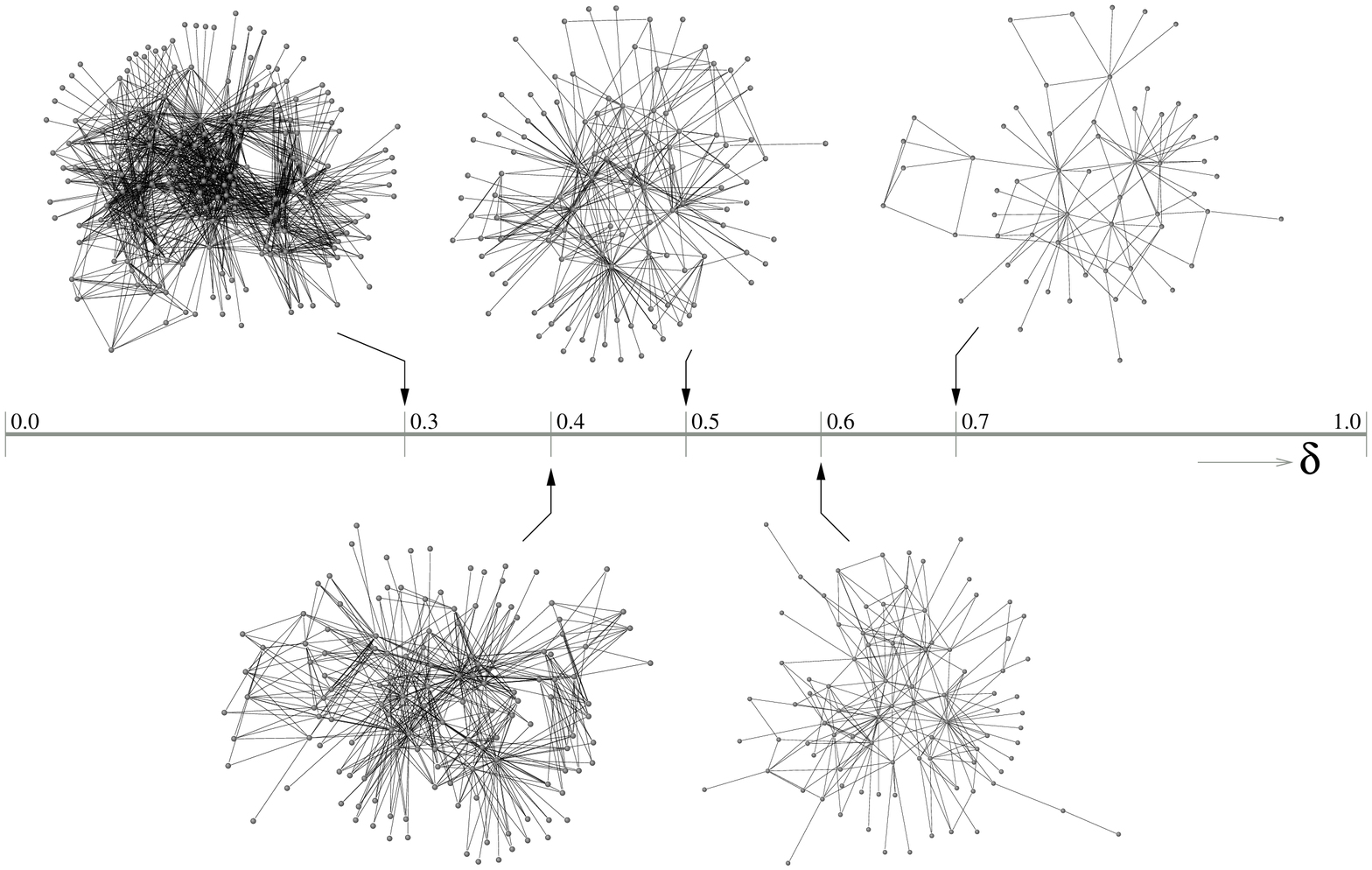}}
  \caption{
    \label{graphsdelta}
The architecture of the proteome map, as generated by the simple model
for different values of the deletion rate $\delta$. As predicted by the
mathematical  model,  two well-defined  phases  are  present. For  the
first, when $\delta<\delta_c=0.5$, the protein map is highly connected
and   most   elements   have   links  to   others.   Conversely,   for
$\delta>\delta_c$ the  graph is fragmented  into many pieces  and many
components have no links or belong to small isolated subnets. Close to
the transition domain, we have a sparse graph with the statistical
features displayed by the real proteome map. Such graph displays modular 
organization, in spite of a complete lack of functionality in 
the definition of the model rules.}
\end{figure*}

The model analysed in  \cite{Sole2002_ACS} is defined by the following
rules. We  start form a  set $m_0$ of  connected nodes, and  each time
step we perform the following operations

\begin{itemize}

\item[(i)] One node of the graph is selected at random and duplicated

\item[(ii)] The links emanating from the newly generated node are
  removed with probability $\delta$. 

\item[(iii)] New  links (not previously present  after the duplication
  step) are created between the new node and all any other node 
with probability $\alpha$. Although available data indicate that new
  interactions are likely to be formed preferentially towards proteins
  with high degree here we do not consider this constraint.

\end{itemize}

Step (i) implements  gene duplication, in which both  the original and
the  replicated proteins  retain the  same structural  properties and,
consequently, the  same set of  interactions. The rewiring  steps (ii)
and  (iii) implement the  possible mutations  of the  replicated gene,
which translate  into the deletion  and addition of  interactions with
different proteins,  respectively. The  process is repeated  until $N$
proteins have been obtained.

The  model  described  in  \cite{Vazquez2003}  is  very  similar,  but
introduces  some relevant  differences. Here  duplication (i)  is also
followed by  two probabilistic rules which  operate independently. The
first (ii) is node deletion. For each of the nodes $p_j$ linked to the
two $p_i$ and its duplicate $p_i'$,  we choose randomly one of the two
links   $\xi_{ji},   \xi_{ji'}$  and   remove   it  with   probability
$\delta$. Additionally, a new  interaction connecting the two proteins
(the  parent  and  the  duplicated)  is  introduced  with  probability
$\pi$. The last  rule will naturally increase the  number of triangles
in  the system  and thus  provide a  source of  high  clustering.  The
rewiring  process seems to  be more  appropriately defined,  since the
removal of  one of  the alternative links  allows to  ``conserve'' the
function  that was somehow  present before  the duplication  event. In
Sole's  model, the  whole  set of  links  of the  duplicated gene  are
preserved and loss of connections  affects only the new copy. By using
Vázquez's approach,  more flexibility  is allowed and  the interaction
map is more likely to remain connected. As defined, it is important to
note that duplicates will diverge only to some extent: if a duplicated
gene  with degree  $k_i$  is  duplicated, only  $\delta  k_i$ will  be
removed on average. To reach higher levels of divergence (as suggested
in the real proteome) we need to remove links from the rest of the map
(and not just from the duplicate). Such a refinement is well based and
has been  also considered  (see discussion) providing  essentially the
same results in relation with  network architecture (Sol\'e et al., in
preparation).

The two models collapse into a single mean field description where the
average connectivity follows the dynamics:
\begin{equation}
 {d K_n \over dn}  = {1 \over n} 
 \left (  K_n + \phi_{\alpha}(n,K_n) - 2 \delta K_n \right )  
\end{equation} 
where     $\phi=2\alpha(n-K_n)$      in     Sol\'e's     model     and
$\phi=2\alpha(n-K_n)=\pi$ in Vázquez's  model. Actually, in a previous
paper \cite{Ric1,Ric2} we showed that  in order to  have convergence in
the system towards a scale-free stationary distribution we need a very
small rate  of link addition  (consistently with observations).  If we
assume  that $\alpha  \sim  O(1/n)$ then  a  single link  is added  on
average  each  step and  thus  the two  models  are  identical in  the 
low-addition limit: specifically, if the graph
is sparse, we have $\alpha(n-K_n) \approx \pi$. In this case we have a
dynamical equation
\begin{equation}
 {d K_n \over dn} +  {2 \delta - 1 \over n} K_n = {2 \pi \over n}  
\end{equation} 
which has an associated general solution:
\begin{equation}
 K_n = e^{-\eta(n)}  
 \left ( 2\pi \int {e^{\eta(n)} \over n} dn + C \right ) 
\end{equation} 
where $\eta(n) = \int  (2 \delta - 1)dn/n =  (2 \delta - 1)\ln n$. 

This gives:
\begin{equation}
 K_n = {2 \pi \over 2 \delta -1 } + 
    \left ( K_0 - {2 \pi \over 2 \delta -1 } \right )  n^{-(2\delta-1)} 
\end{equation} 
if  $\delta>\delta_c=1/2$, the  previous system  converges to  a graph
with a finite average degree
\begin{equation}
 K_{\infty} = \lim_{n\to\infty} K_n  = {2 \pi \over 2 \delta -1 } 
\end{equation} 
Otherwise,  the average connectivity  will be  $K_{\infty} \rightarrow
\infty$.   The critical  removal  rate $\delta_c=1/2$  thus defines  a
phase  transition separating  a phase  with a  highly-connected system
($\delta<\delta_c=1/2$) from a  sparse phase ($\delta>\delta_c$) where
a finite number of links will  be observed. At this phase, the network
becomes fragmented into  many pieces. It is interesting  to note that,
under the  present conditions, the  long-term behavior of  the average
connectivity does  not depend on the  rate of link  addition.  What is
really important  is that the rate  of link addition  and link removal
are similar, so that $\langle k \rangle$ can reach a stationary
value.  Moreover,   it  can  be   shown  that  although   no  explicit
preferential attachment is included here, the multiplicative nature of
the process  (in which proteins having  more links are  more likely to
have them  copied) actually leads  to an {\em  effective} preferential
attachment \cite{VazquezPRE2003}.

\begin{figure}
  {\centering \includegraphics[width=8cm]{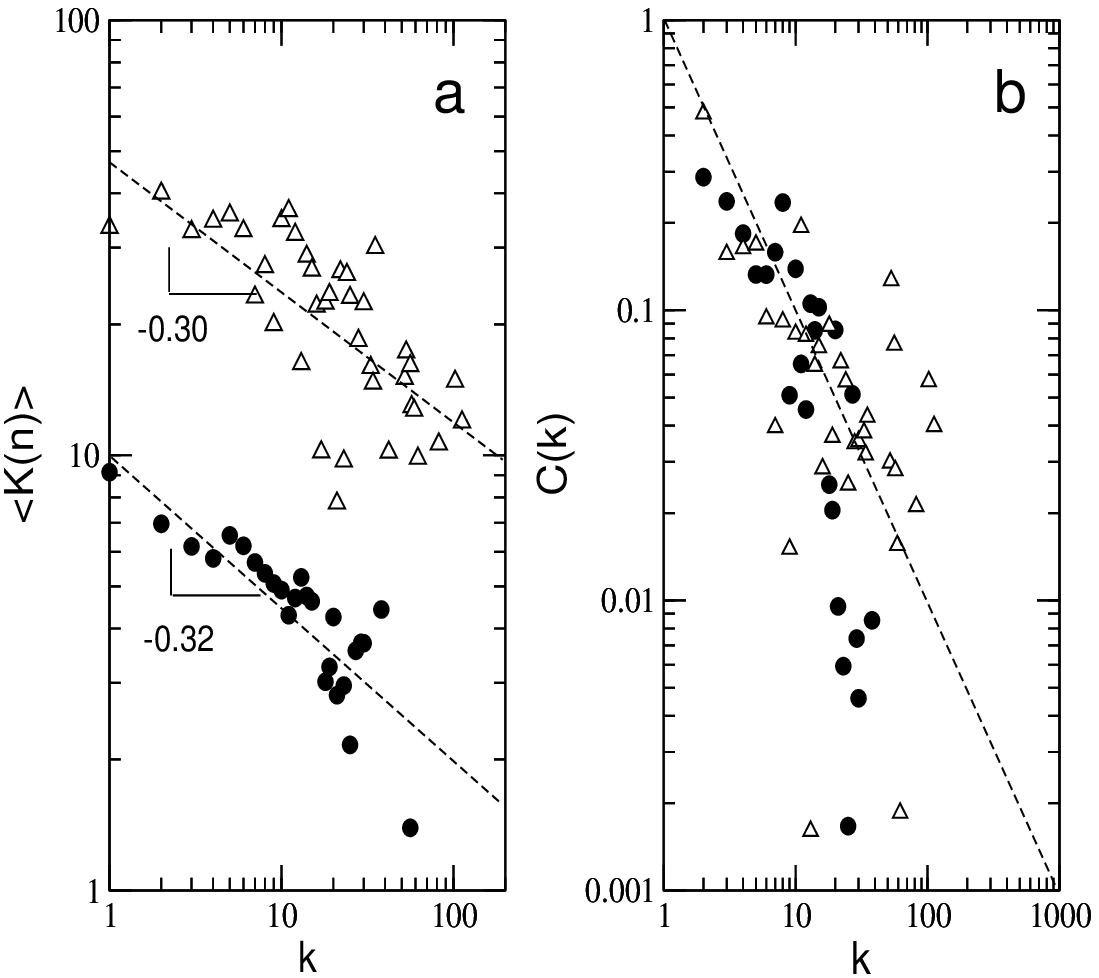}}
  \caption{
    \label{clustcorrel}
Comparison between correlations in the proteome model at $\delta=0.55$
and the observations form the yeast proteome map (here the models used
provided  a connected  components with  the same  size than  the yeast
map). In (a)  the correlation scaling for the  average connectivity is
shown,  with   a  fit  for   both  yeast  data  (circles)   and  model
(triangles).  Although the scaling  behavior is  the same,  the larger
number of  links predicted  by the model  shifts the  expected average
towards higher  values.  In  (b) we plot  the scaling behavior  of the
clustering  against degree.   The dashed  line indicates  the expected
scaling  behavior  assuming  hierarchical  organization  (see  text).}
\vspace{0.25cm}
\end{figure}

We  can test this  prediction by  studying the  behavior of  the model
under different rates of link deletion. In order to measure the impact
of  this rate  on network's  architecture, we  use two  different, but
closely related  measures: (1)  the normalized largest  component size
$S$ and (2) the average, normalized component size $\langle s\rangle$.
If ${\cal C}(\Omega)=\{\Omega_1,\Omega_2, ..., \Omega_c \}$ is the set
of connected components (subgraphs) of the proteome map, so that
\begin{equation}
\Omega=\bigcup_{i=1}^c \Omega_i
\end{equation}
and $n_i=\vert \Omega_i \vert$ indicates their 
size (so that $\sum_i n_i = N$), we define:
\begin{equation}
S = {1 \over N} \max \{ n_i \} 
\end{equation} 
\begin{equation}
\langle s\rangle = {1 \over N} \left ( {1 \over c} \sum_{i=1}^c  n_i \right ) 
\end{equation} 
In figure  3(a-b) we display the  two measures against  $\delta$ for a
$N=10^3$  protein network.  Close  to $\delta_c$  we can  appreciate a
clear  change.   The  two  phases  are clearly  identified,  with  the
connected one showing $S\approx 1, \langle s\rangle \approx 1$ and the
fragmented  phase  showing   $S\approx  1/N,  \langle  s\rangle\approx
1/N$.  In  3(a)  we  can  see  that  $S$  decreases  slowly  close  to
$\delta_c$, where only about half of the nodes remain connected within
the largest  component. The sharpness  of the transition  becomes much
more  obvious  in 3(b).  Here  we  clearly  appreciate the  impact  of
rewiring on  network's structure, indicating that a  large fraction of
the   overall  network   structure  is   formed  by   small,  isolated
components.In  figure  4  we  can  see some  examples  of  the  graphs
generated  (largest   components)  obtained  at   different  rates  of
deletion.

\section{Hierarchical organization, modularity and correlations}

Previous papers on  genome/proteome architecture have mainly described
the heterogeneous  character of the  protein-protein map as well  as a
few  large-scale  features  as   the  clustering  coefficient  or  the
network's  diameter.   Beyond  such  measures, which  only  contain  a
limited part of network's  structure, correlations offer a much better
view of their internal organization.

One measure of  correlations can be easily obtained  by looking at the
set  of  conditional probabilities  $p_c(k\vert  k')$  that a  protein
having  $k$  links   is  connected  to  a  protein   with  $k'$  links
\cite{MaslovSneppen}. If no correlations exist (as it would occur in a
purely random  network) then we  would have $p_c(k\vert  k')=p(k)$. We
can analyse the average connectivity $\langle k(n)\rangle$ defined as:
\begin{equation}
\langle K(n) \rangle = \sum_{k'} k' p_c(k\vert k') 
\end{equation} 
(which is  just $\langle k(n)\rangle=\langle k\rangle$  in the absence
of  correlations).  Data  from   PIN  gives  a  scaling  law  $\langle
k(n)\rangle \sim k^{-\nu}$ with $\nu  \approx 0.30 \pm 0.03$, as shown
in figure 5(a) (open triangles).  The result from Sol\'e's model close
to the phase transition is  also shown (black circles), with $\nu_{SM}
\approx  0.32 \pm  0.06$.  This  scaling law  indicates that  there is
strong  anticorrelation among nodes  with low  and high  degree.  Hubs
tend to be unconnected among them, and instead they are connected with
low-degree  proteins. This  type  of  network is  also  known as  {\em
disassortative}.  The scaling  appears to behave the same  way in both
data and model,  but the higher average connectivity  predicted by the
model actually shifts  the {\em in silico} law  towards higher values.
This difference is easily removed  when the model is expanded allowing
to remove  links in  a correlated way  not restricted to  the recently
duplicated node.

Similarly,   the  presence   of  hierarchical   organization   can  be
highlighted by  looking at  the clustering-degree function  $C(k)$. As
discussed in  the introduction,  this function provides  a statistical
test for the presence of hierarchies in graph structure. As we can see
in  figure  5(b),  both the  proteome  map  and  its {\em  in  silico}
counterpart display a non-uniform  behaviour of the clustering against
degree.  This  gives  further  support  to  the  presence  of  modular
structure (see below).

A more  detailed, complete view  of the correlation structure  of both
model and real  maps is given by correlation  profiles (CP) as defined
in  \cite{MaslovSneppen}.  In order  to compute  it, we  calculate the
joint probability $P(k_i,k_j)$  with $1 \le k_i, k_j  \le N$, that two
proteins are connected to each  other. We also compute the probability
$P_r(k_i,k_j)$ obtained by randomizing the  same network (i. e. a null
model with no correlations). Significant correlations will be observed
through systematic deviations of the ration
\begin{equation}
\Gamma(k_i, k_j) = { P(k_i,k_j)  \over P_r(k_i,k_j)  }
\end{equation} 
from the null  model (i. e. deviations from  $\Gamma(k_i, k_j)=1$). In
figure  6 the  results  from the  CP  are shown  for  both real  yeast
proteome (a) and different models (b-d).

Two prominent features are  observed in 6(Y).  The first, consistently
with the previous analysis of  $\langle k(n) \rangle$, is the presence
of anticorrelation  between nodes of given degree.   This is indicated
by the red spots: nodes with high degree are not connected among them,
but typically linked to proteins  with low degree. A second feature is
the presence  of significant  correlation among proteins  with degrees
close  to  $k_i \sim  10$.   Actually, a  wider  domain  close to  the
diagonal is  implicated, indicating the  presence of sets  of proteins
forming  multiprotein complexes  \cite{MaslovSneppen}. Both  DD models
(figures  6(A,B), here  (A) Sol\'e's  model and  (B)  Vazquez's model)
naturally  give the  red spots  at the  correct locations  in  the CP.
Additional correlations  are shown near the $(k_i,  k_k) \sim (10,10)$
zone.  In (B) two spots  are observed around this location, whereas in
(C) the correlation is present close to the diagonal but although less
pronounced. The  first feature is  a result of the  intrinsic dynamics
shown by  the DD models,  in which rapid divergence  after duplication
allows  initially  linked hubs  to  become  disconnected.  The  second
feature  provides a good  example of  how truly  functional constrains
(those defined  by protein complexes) shape  real genome architecture.
As discussed by Maslov and Sneppen, multiprotein complexes are largely
responsible  for this  feature. The  fact that  the DD  models  do not
display this structure is an indication that the lack of functionality
is likely to explain the lack of the observed pattern.

\begin{figure}
  {\centering \includegraphics[width=8cm]{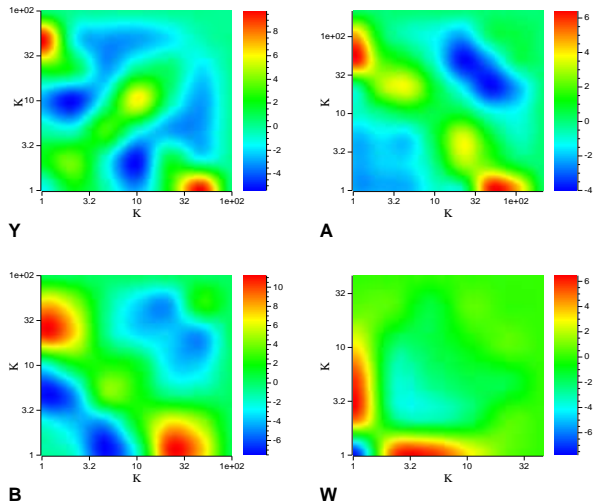}}
  \caption{
    \label{maslov4maps}
     Comparison between  degree correlations in the  proteome model at
$\delta=0.55$ and  the observations form  the yeast proteome  map. (Y)
Real  yeast network,  (A) Solé's  model, (B)  Vázquez's model  and (W)
Wagner's model.   In (Y), one can  observe the two red  spots for high
degree nodes that are linked  to low degree ones, and some correlation
at about  $k\approx 10$. As it  is apparent, (A) and  (B) resemble the
real case  (Y), whereas (W)  does not, highlighting the  importance of
the  internal  structure besides the degree distribution.  }
\vspace{0.25cm}
\end{figure}

For comparison, we also  display the correlation profile obtained from
a different  model of proteome evolution  \cite{WagnerA2003}.  This is
actually a particular example of  a model presented by Dorogovtsev and
Mendes, \cite{DorogovtsevBook}  (DM) in which no change  in the number
of nodes is allowed, only rewiring. Here duplicated genes play no role
and thus no correlations  from duplication are preserved. Interactions
are added  and eliminated at  given rates, being these  rewiring rules
applied using preferential attachment.  Under a strict balance between
addition  and deletion  (again,  we have  a  phase transition  between
explosion and fragmentation) a power law in the degree distribution is
obtained.  But  any correlation  is lost under  this type  of approach
(such as the lack of clustering or modularity). This is illustrated in
figure 6(d) where  the correlation profile obtained from  the DM model
parameters used in  \cite{WagnerSFI2002,WagnerA2003} is shown. A
visual inspection reveals a proteome map with little relation with the
observed one. This results should prevent us of performing comparisons
between model  and real network  data limited to a  single topological
property.

The previous correlations displayed by DD models and the evidence of a
hierarchical organization strongly indicate  that some type of modular
architecture should be expected.  In order to properly detect modules,
we use  the topological overlap method  \cite{Ravasz2002}.  An overlap
matrix $O_T(i,j)$ is defined as:
\begin{equation}
O_T(i,j) = {J_n(i,j) \over \min\{k_i, k_j\}}
\end{equation}
Here  $J_n(i,j)$ is the  number of  proteins to  which both  $p_i$ and
$p_j$ are  linked.  The denominator  gives the smallest degree  of the
pair $\{k_i, k_j\}$. Since both  terms are constrained to the interval
$(0,N)$, the overlap matrix is  normalized, i. e.  $0 \le O_T(i,j) \le
1$.  This matrix can be  then displayed as a two-dimensional plot with
a color  scale indicating the  relative amount of overlap  between two
given nodes.   The set of nodes  is also arranged  with an appropriate
algorithm so that  elements belonging to the same  module appear close
within the matrix.  Two examples of these maps are  shown in figure 7,
for  the two  models  explored  here. We  can  clearly appreciate  the
presence  of  proto-modules,  as  shown  by the  clusters  of  closely
connected elements.  A hierarchy  of such clusters,  defining a  set of
nested modular structures, is observed.

\begin{figure}
  {\centering \includegraphics[width=8cm]{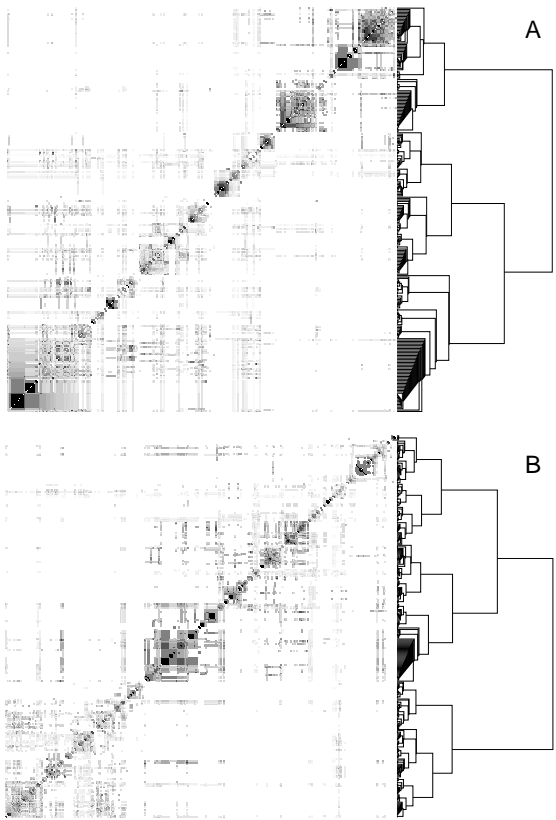}}
  \caption{
    \label{solevazquez}
Topological overlap  matrix from the two DD models considered here. (A)
Sole's  model; (B) Vázquez's  model.  The  modular architecture  of the
interaction maps has been obtained close to the phase transition point
(here $\delta=0.5$)}
\end{figure}

\section{Discussion}

The emergence of modularity is one of the key problems of evolutionary
biology.  Modules  are common to  both natural and  artificial systems
\cite{Hartwell1999}  and it  is generally  agreed  that modularization
allows a  well-defined functional separation  with enhanced robustness
against  component failure. One  should expect  to observe  modules as
slowly emerging  from small subgraphs performing  some functional role
(such allowing  bistability, or the  creation of stripes in  the early
embryo) and adding more components  able to tune their performance and
increase their adaptability and  robustness. In this way, compartments
performing specialized  functions should  be expected to  emerge. This
might  have been  the  case of  the  evolution of  ontogeny in  neural
circuitry:  a process of  {\em parcellation}  would have  been shaping
neural  structures  through  a  mechanism  involving  segregation  and
isolation. It  is actually  interesting to  note that
such a  parcellation process deals with two  essential components: the
presence  of  some   redundancy  in  cell-cell  (neural)  interactions
followed by loss of one or more  inputs to a cell.  In other words, we
need first  to have several  neighboring neurons, likely to  have been
obtained  from cell  duplication  of a  common  parental strain.   The
initial set of  neurons will be more densely  connected and afterwards,
specialization  will  occur  by  loosing  some  links.   This  process
strongly reminds us the one taking place in the proteome map, although
some fundamental differences are also present.

The  proteome  model provides  a  surprising  counterexample of  these
intuition. Here  local rules  are able to  shape some key  features of
global structure.  Such as  scenario seems to  be rather  general, and
might  have  implications  for  the  origins of  metabolic  paths  too
(Lehmann,  Ravasz and  Wuchty,  submitted paper).   Instead of  slowly
creating modules  from significantly rewiring sub-parts  of the graph,
modules appear to  be present as a consequence of  the DD process.  As
illustrated by the previous figure, proto-modules spontaneously emerge
and are thus a pre-pattern. Such a pre-defined structure could then be
used  in order  to perform  cellular functions.  It is  interesting to
compare  these  structures with  those  present  in technology  graphs
\cite{Ferrer2001a,Valverde2002}.

What can be learn in general  from this example? On the one hand, this
study provides an example of modularity ``for free'': there is no need
of natural selection fine-tuning the system in order to obtain a large
amount of correlations.   Close to the narrow domain  of high deletion
rates  scale-free  architecture emerges  in  a  natural  way.  Such  a
conjecture agrees  with the  view of evolution  as constrained  and to
some extent  shaped by emergent  properties \cite{Kauffman1993,Signs}.
But several  relevant questions  emerge. One deals  with the  rates of
link  addition and  removal. Why  are  we observing  these high  rates
leading to a sparse graph?  Two main possibilities emerge.  One has to
do with  the requirement of a  sparse graph in order  to avoid dynamic
instabilities. Specifically,  if the activity of the  network is taken
into account, positive and negative links between different parts of a
regulatory  network  can trigger  the  emergence  of chaotic  dynamics
\cite{SoleSatorras2002}.   Feedback loops in  particular are  known to
destabilize  complex networks and  a sparse  graph would  easily avoid
them  to break  system's  stability.  By  tuning  the average  degree,
selection   might  have   reached  a   stable,  robust   network  with
proto-modules embedded within its  basic architecture. Another is that
such  proto-modules might  have been  the real  target of  selecting a
sparse graph. Modules themselves isolate different parts of the system
and   thus   a   mechanism   favoring  their   emergence   (even   as
proto-structures)  might  have   been  successfully  chosen.   Further
studies should  consider these possibilities by  exploring the internal
organization of the protomodules, to be compared with the one observed
in real maps.

\begin{acknowledgments}
  The authors would  like to thank the members  of the Complex Systems 
 Lab for useful discussions.  This work was supported by a 
grant  BFM2001-2154 (RVS), the Generalitat  de Catalunya  (PFD, 
2001FI/00732) and The Santa Fe Institute.
\end{acknowledgments}

\vspace{0.6cm}

\bibliographystyle{plain}
\bibliography{networks,bionetworks,evodevo,general,koonin}

\end{document}